\documentclass{ws-procs975x65}

\begin{document}

\title{\uppercase{The evolution of non-adiabatic pressure perturbations during
multi-field inflation}}

\author{IAN HUSTON}
\address{Astronomy Unit, School of Physics and Astronomy, Queen Mary University
of London, \\
Mile End Road, London, E1 4NS, United Kingdom}

\author{ADAM J. CHRISTOPHERSON}
\address{School of Physics and Astronomy, University of Nottingham, \\
University Park, Nottingham, NG7 2RD, United Kingdom}

\begin{abstract}
Non-adiabatic pressure perturbations naturally occur in models of
inflation consisting of more than one scalar field.
The amount of non-adiabatic pressure present at the end of inflation
can have observational consequences through changes in the curvature
perturbation, the generation of
vorticity and subsequently the sourcing of B-mode polarisation.
In this work, based on a presentation at the 13th Marcel Grossmann Meeting, we
give a very brief overview of non-adiabatic pressure perturbations in
multi-field
inflationary models and describe our recent calculation of the
spectrum of isocurvature perturbations generated at the end of
inflation for different models which have two scalar fields.
\end{abstract}


\bodymatter

\section{Introduction}
In part due to the possibility of observable non-Gaussian signals, the
study of inflationary scenarios has recently focussed on models
with more than one operative scalar field. The extra degrees of freedom
present in these physical systems support isocurvature or non-adiabatic pressure
modes in addition to the usual adiabatic mode.

Cosmological perturbation theory is a well-established technique for
analysing the inflationary perturbations which allows us to numerically compute
the non-adiabatic pressure throughout the inflationary era.
(See Ref.~\refcite{MW2008} and references therein for a
comprehensive introduction to perturbation theory.)

One reason for considering the non-adiabatic pressure
perturbation produced towards the end of inflation is that non-adiabatic
pressure perturbations can source vorticity at second order in perturbation
theory \cite{vorticity}.
This vorticity is expected to have some effect on the CMB, as vector
perturbations can source B-mode polarisation.

In addition the presence of non-adiabatic pressure at the end of inflation can
affect the predictive power of an inflationary model. The conservation of the
curvature perturbation at large scales requires the absence of
non-adiabatic pressure\cite{MW2008}. To connect inflationary results with
CMB data in the presence of non-adiabatic modes then requires that the full
evolution through reheating and radiation domination is known.

\section{Non-adiabatic pressure in multi-field inflation}
By considering scalar perturbations at linear order to the energy-momentum
tensor and a flat Friedmann-Robertson-Walker metric we can find the equations
of motion of the system using the perturbed Einstein equations. Here we
consider perturbations in the uniform curvature gauge where the scalar curvature
perturbation is identically zero. In the following we show results for a system
of two scalar fields $\varphi$ and $\chi$, but the results are generalisable to
many field systems and the numerical code discussed below handles the full
multi-field case.

In general a fluid, in this case the combined scalar field fluid, is not purely
adiabatic. The full
pressure perturbation $\delta P$ can then be split as 
\begin{equation} 
\label{eq:dPsplit}
\delta P = \delta P_{\rm nad} + c_{\rm s}^2\delta\rho\,,
\end{equation}
where $\delta P_{\rm nad}$ is the non-adiabatic pressure perturbation,
$c_\mathrm{s}^2 =
\dot{P}/\dot{\rho}$ is the adiabatic sound speed for the fluid and $\delta
\rho$ is the energy density perturbation.

For the case of two scalar fields with a combined potential $V(\varphi, \chi)$,
the non-adiabatic pressure perturbation at linear order is given by
\begin{align}
\delta P_{\rm nad} = &\frac{8\pi
G}{3H^2}(V_{,\varphi}\dot{\varphi}+V_{,\chi}\dot{\chi})(\dot{\varphi}
\delta\varphi+\dot { \chi } \delta\chi)
-2(V_{,\varphi}\delta\varphi+V_{,\chi}\delta\chi)\\
&-\frac{2}{3H}\frac{(V_{,\varphi}\dot{\varphi}+V_{,\chi}\dot{\chi})}{(\dot{
\varphi } ^2+\dot { \chi}^2)}
\Big[\dot{\varphi}\dot{\delta\varphi}+\dot{\chi}\dot{\delta\chi}+V_{,\varphi}
\delta\varphi+V_ { , \chi}\delta\chi\Big]\,,\nonumber
\end{align}
where $V_{,\varphi} = \partial V/\partial \varphi$ and
$\delta\varphi$ and $\delta\chi$ are the linear perturbations in the $\varphi$
and $\chi$ fields.

\section{Discussion}
In Ref.~\refcite{Huston:2011fr} we numerically calculated the non-adiabatic
pressure for a number of interesting inflationary models. We used the Pyflation
package for Python, an open
source numerical package created by one of the authors and described in
Refs.~\refcite{hustonmalik2,hustonmalik,pyflation}. The numerical
system evolves the
background and first order perturbations of the scalar fields from deep inside
the cosmological horizon until the end of inflation. For $n$ scalar fields the
full $n\times n$ matrix of mode amplitudes for the quantum creation operators
are evolved and only one run is necessary. This is in contrast to other
approaches in which $n$ separate runs are required, with the initial condition
for only one of the field perturbations being non-zero in each run.

\begin{figure}
 \centering
 \psfig{width=0.75\textwidth, file=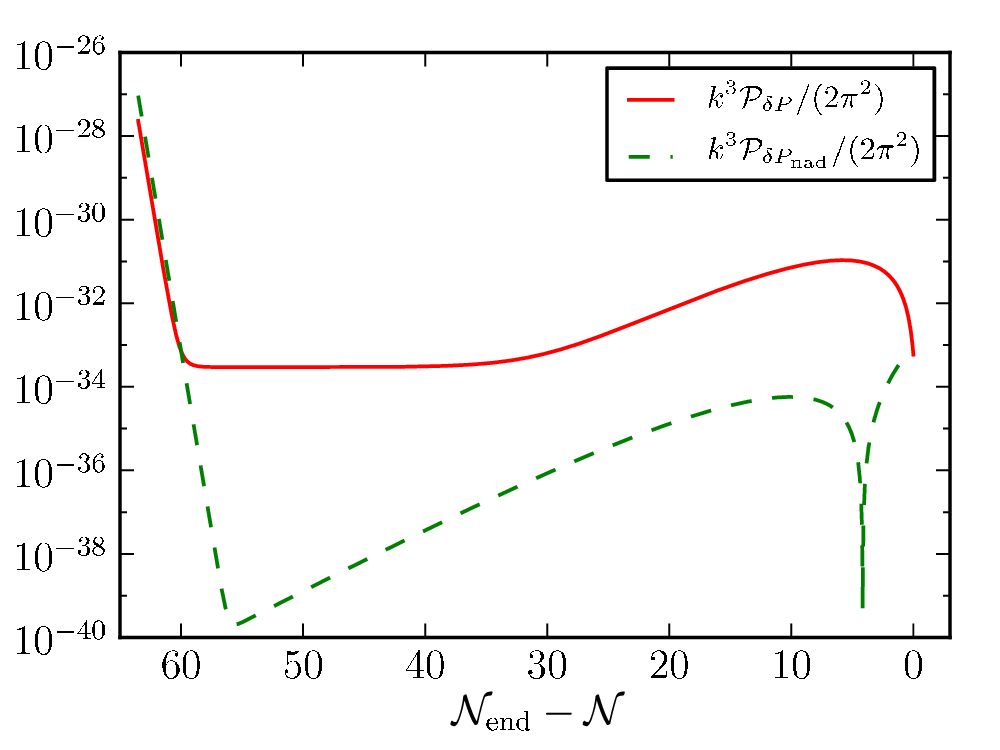}
 \caption{A comparison of the scaled power spectra 
of $\delta P$ (red straight line) and $\delta P_\mathrm{nad}$ (green dashed
line)
for the product exponential
potential at the {\sc Wmap} pivot scale. The x-axis denotes the number of
e-foldings before the end of inflation, $\mathcal{N}_\mathrm{end}$.}
 \label{fig:prodexp-dPnaddP_N}
\end{figure}

We examined the production of non-adiabatic
pressure in Ref.~\refcite{Huston:2011fr}  for three potentials widely considered
in the literature: a double quadratic model
$\frac{1}{2}m_\varphi^2\varphi^2+\frac{1}{2}m_\chi^2\chi^2$,
a quartic model $\Lambda^4[
(1-\chi^2/v^2)^2+\varphi^2/\mu^2 + 2\varphi^2\chi^2 /(\varphi_{ \rm
c}^2v^2)]$, and a product exponential model $V=V_0 \varphi^2 e^{-\lambda
\chi^2}$. 
The results for the non-adiabatic pressure of the product exponential model are
shown in Fig.~\ref{fig:prodexp-dPnaddP_N}. In
calculating these results we have set $\lambda = 0.05 /M_{\rm pl}^2$ and
normalized the power spectrum to the WMAP value by setting
$V_0 =5.37\times 10^{-13} M_{\rm pl}^2$. The initial field
values are $\varphi_0=18M_{\rm pl}$ and $\chi_0=0.001 M_{\rm pl}$.

These results are for one particular wavemode, the WMAP pivot scale, and compare
the evolution of the full pressure perturbation with the non-adiabatic part over
the course of the inflationary expansion. For this model the amount of
isocurvature present at the end of inflation is significant enough to warrant
careful examination of the evolution after inflation, especially during the
reheating phase. As such any prediction of the statistics of curvature
perturbations given at the end of inflation is subject to modification during a
reheating phase and radiation domination. We will explore the role of
non-adiabatic pressure and the evolution of curvature perturbations during
reheating in a future publication\cite{Huston:2012}.

\section*{Acknowledgements}
IH is supported by the STFC under Grant
ST/G002150/1, and AJC is funded by the Sir Norman Lockyer
Fellowship of the Royal Astronomical Society.

\bibliographystyle{ws-procs975x65}
\bibliography{isocurvpapers}

\end{document}